# Semi-Automatic Terminology Ontology Learning Based on Topic Modeling

Monika Rani[*], Amit Kumar Dhar and O. P. Vyas

*Department of Information Technology,
Indian Institute of Information Technology, Allahabad, India
monikarani1988@gmail.com*

Abstract- Ontologies provide features like a common vocabulary, reusability, machine-readable content, and also allows for semantic search, facilitate agent interaction and ordering & structuring of knowledge for the Semantic Web (Web 3.0) application. However, the challenge in ontology engineering is automatic learning, i.e., the there is still a lack of fully automatic approach from a text corpus or dataset of various topics to form ontology using machine learning techniques. In this paper, two topic modeling algorithms are explored, namely LSI & SVD and Mr.LDA for learning topic ontology. The objective is to determine the statistical relationship between document and terms to build a topic ontology and ontology graph with minimum human intervention. Experimental analysis on building a topic ontology and semantic retrieving corresponding topic ontology for the user's query demonstrating the effectiveness of the proposed approach.

## 1. Introduction

Web 2.0 allow a user to participate in information sharing by writing reviews, comments, and feedback on sites not like the Traditional Web (web 1.0: 1990-2000) where user are passive information receivers. But there are certainly significant problems in the Web 2.0 (2000-2010) which lead to the birth of Web 3.0 (2010-2020). For example, in Web 2.0 (Web of Document) content is not machine-readable format. Information on the web is in a heterogeneous source format (HTML, XML, etc.) and thus can't be used for purpose of integration, analysis, and intelligent data analysis. Tim Berners-Lee (Berners-Lee et al., 2001) gave a vision of Semantic Web (Web 3.0) which can connect data semantically in machine-readable format, provide common vocabulary using ontologies. Basically, Web 3.0 is an extension of Web 2.0 which aim to provide a common framework for sharing data across application boundaries. Web 3.0 also called the Web of Data (WoD) whereas Web 2.0 was known as the web of documents. We are moving towards the era of Web of Documents to Web of Data. Web 2.0 perform syntactic search whereas web 3.0 capable of doing the semantic search using semantic web and ontologies technology. Web of Data (Web 3.0) provides one more level deeper information thus it can answer more complex queries. Also, Web 3.0 allow ordering and structuring of content, provide machine-readability, content reusability and facilitate agent interaction on the Web using ontologies (Gruber, 1993). Ontologies are the pillar of semantic web used for knowledge representation. Ontology learning (OL) greatly helps ontology engineers for building their own ontologies for a particular domain. The steps for OL are Import & reuse, Extract, Prune, Refine and Validation (Maedche, 2012). Different approaches for OL can be classified based on the type of knowledge resources for which to learn ontology (structured, semi-structured, unstructured format of data), Level of automation (semi or full automated), Learning target (concepts, Taxonomy, Conceptual relation, Attributes, Instances, Axioms), purpose (creating/updating), Learning techniques (Linguistics, Statistical and Hybrid approaches) (Al-Arfaj & Al-Salman, 2015).

The present research focuses on Ontology Learning (OL) disabilities such as: automatic conversion of Text to Ontology (Text to Onto) (Wong et al., 2012), Never Ending Language Learning (NELL) (Carlson et al., 2010) as data is continuously increasing on the web and database, Open Information Extraction (OIE) etc. So, in continuation of that our main focus is to learn terminology ontology based on topic modeling and then provide primarily step for semantic-based query retrieval (Topics and Words Detection). In this paper, two topic modeling algorithm are explored namely: Latent Semantic Indexing (LSI) & Singular Value Decomposition (SVD) and MapReduce Latent Dirichlet Allocation (Mr.LDA) for learning terminology ontology. Our target is to achieve terminology ontology from textual data using topic modeling and primarily provide semantic-based query retrieval for Topic and Word Detection for a domain. Though there are powerful search engines like Google, Bing, Yahoo!, etc. but they are limited in their capability of retrieving a relevant list of the Web of Documents (WoD) rather than required data. Also, unfortunately, most of the textual data available on the internet are not machine-readable format. It is also quite cumbersome to organize and access as textual data which is continuously growing per second. Existing search engine as retrieval result brings out as the Web of Document (WoD) where as if we classify the large corpus into ontology then search required content (topic and word) which is the web of data (Web 3.0) is time-saving and cost of searching. Therefore, our approach is to build semi- automatic terminology ontology using topic modeling algorithm (LSI & SVD and Mr.LDA) to classify topics and associated words for knowledge management and semantic retrieval.

Ontology learned can be used in various fields as they have the ability to reduce communication gap by providing the common vocabulary, in association with agents provide a personal assistant for real-time applications also classify and recommend content for the user. The proposed terminology ontology is based on topic modeling (Mr.LDA model) can be used in other fields of research examples: Semantic Retrieval (Tran et al., 2007), Semantic Web in E-learning (Rani et al., 2015), Data Enrichment & Mining (Dou et al., 2015; Abedjan, 2014), Topic Detection and Tracking, Natural

Language Processing (NLP), Semantic search over heterogeneous networks (Tang et al., 2011; Hogan, 2011), Knowledge Engineering and Management, Electronic Commerce, Sentimental Analysis, Scientific Exchange, Bio-informatics, Biomedical, Human Computer Interaction. Also, build ontology is beneficial for Ontology-based Association Rule Mining (ARM), Classification, Clustering, Link Prediction, Information Retrieval, Recommendation System, etc.

The rest of paper is organized as follows. Section 2 describes Ontology Learning (OL), Latent Semantic Indexing (LSI), probabilistic Latent Semantic Indexing (pLSI), MapReduce Latent Dirichlet Allocation (Mr.LDA). Section 3 introduces a general description of the explored topic modeling (LSI & SVD and Mr.LDA). Experimental results and conclusion is reported in Sections 4 and 5 respectively.

## 2. Review of related works

Topic Modeling is a form of text mining, where we can retrieve required text from the large corpus. Topic Modeling uses various algorithms or a modeling approach to organizing, summarize large corpus and retrieve require text. In this section, some related works, namely those regarding Ontology, Topic Models (Latent Semantic Indexing (LSI) & Singular Value Decomposition (SVD) and Latent Dirichlet Allocation (LDA)), are briefly reviewed.

### 2.1 *Ontology*

Ontology is the study of semantics, existing in the world, which can be formally defined as, a formal and explicit specification of a shared conceptualization (Gruber, 1995). It means ontology explicitly define the rich relation between concepts which are machine readable and sharable among a group of people. Since the introduction of Web 3.0, there has been an exponential rise in the amount of data that is accumulated each day. This data has to be well defined and explicitly represented so that it can be shared and used by humans and machines, enabling them to work together in a better way.

2.1.1 *An Ontology generally consists of:*

- Individuals (aka instances): consider every existing object, example: you and me.
- Concept (aka classes): a group of existing objects example: person, organization, etc.
- Attributes: describe the property of concept example: height, weight, etc.
- Relationships: represent the relation by which two concepts are associated, example: students are affiliated with IIIT-Allahabad College.

2.1.2 *Following are the advantages of making an ontology:*

- An ontology provides us with a common vocabulary for a domain (Cakula and Salem, 2013).
- Merge and expansion of ontologies based on their metadata are easy.
- An ontology defines content unambiguously.
- To separate operational knowledge from domain knowledge.
- The ontology allows re-use of content represented in it (Jovanović et al., 2007).
- Ontology provides ordering and structuring of the content store in it (Dzemydiene and Tankeleviciene, 2008).

- A rule can be added to ontology to infer new knowledge.
- Integrate content from a heterogeneous source.
- Effective information sharing, storage, and retrieval of content (text corpus).
- Agent interaction to share content store in an ontology (Jekjantuk and Hasan, 2007).

2.1.3 *Ontology Classification: by level of generality:*

Guarino categorized ontology according to the level of generality (Guarino, 1998):
- Top-level ontology: Its main concerns on general concepts like time, event, action, matter, etc.
- Domain ontology: It provides a common vocabulary to a domain, so various domain knowledge can be interpreted and exchange. An example of domain ontology: Music ontology, Food ontology, Geo ontology, Gene ontology, etc.
- Task ontology: It is based on activity or task example selling or diagnosing.
- Application ontology: Ontologies are built for a specific purpose to share knowledge modeling among various domains.

2.1.4 *Ontology Classification: by level of formality:*

- Informal ontology– It is a taxonomy, example: web directories (Yahoo! Directory), glossary directory, etc. Ontology is rich in term of expressing the relationship, than taxonomy. Taxonomy represents only a hierarchical arrangement of the group.
- Formal ontology– To build a formal ontology, OWL formal language is considered like OWL 1 and OWL 2. Example Cyc and DOLCE.
- Semi-formal ontology– schema structure is considered semi-formal ontology as RDFS language is used rather than a formal language like OWL.

2.1.5 *Ontology can be categories on the basis of purpose into Classification Ontologies and Descriptive ontologies:*

- Classification Ontologies: The document stores are huge and increase with the time on the internet. To classify this document on the basis of relations between terms appropriate hierarchy is used. Searching of a document using the title, subject, and an author can be easily done using classification ontologies.
- Descriptive ontologies: It is the study of being existing, i.e., it can be used to describe all the real world existing entities. Ontology used to describe the Domain, Entity, Relation, and Attribute (DERA) (Giunchiglia et al., 2013).

2.1.6 *Steps for building Descriptive ontologies:*

- Identification of the atomic concepts
- Analysis
- Synthesis
- Standardization
- Ordering and Formalization.

2.1.7 *Current issues in building Ontology:*

- Ontological Engineering refers to the set of activities that concern the ontology development process, the ontology life cycle, the methods and methodologies for building ontologies, and the tool suites and languages that support them (Corcho et al., 2006).
- Ontology learning is the process of converting text to ontology (Text-to-Onto) (Maedche and Staab, 2004).
- Ontology learning from Text to Onto considers Never Ending Language Learning (NELL) as a problem continuously increasing data on the web for this building ontology will be a difficult and ongoing process.

2.1.8 *Ontology operations:*

- Ontology Mapping: It identifies the correspondence between entities of ontologies (Gašević and Hatala, 2006). E-Course content stored in the ontology can be mapped to an answer to learn query. E-learning is one of the application fields of ontology.
- Ontology Matching: It semi-automatically identifies the correspondence between entities of ontology for the purpose of merging and question answering system (Shvaiko and Euzenat, 2013).
- Ontology Merging: This operation merges the entire (or some) heterogeneous ontology source to form a new ontology (De Bruijn et al., 2006).

2.1.9 *Ontology language and the generation of ontology:*

- Familiar ontology languages are OIL, DAML+OIL, XOL, SHOE, and OWL. Interoperability is ensured in the web environment, as these languages are based on the web standards RDF and XML.
- Extended Markup Language (XML): XML (Bray et al., 1998) represents information for both human and a machine readable format drawback of XML is that it represents information in the Syntactic approach rather than Semantic approach.
- Ontology Interface Layer (OIL): OIL uses description logic and XML /RDF for semantic and syntax representation, respectively (Fensel et al., 2000).
- DAML+OIL: Informally RDF Triplets can use to represent DAML, OIL syntax. Basically, DAML OIL gives semantic meaning to RDF Triplet (Horrocks and Harmelen, 2001).
- An XML-based Ontology Exchange Language (XOL): XOL (Karp et al., 1999) is used to exchange ontology definition based on XML.
- Simple HTML Ontology Language (SHOE): SHOE (Heflin et al., 2001) is the first language which allows HTML page to give semantic meaning rather than only syntax analysis.
- Resource Description Framework (RDF): RDF is a W3C standard, used to represent web resource. RDF is a framework used to represent triple which consists of a subject, an object, and a predicate (Lassila and Swick, 1999).
- The Web Ontology Language (OWL): In 2004 W3C recommends Web Ontology Language (OWL) (Dean et al., 2002) as it is more expressive and can represent cardinality and rich relations which RDFS is unable to represent. The limitation of

OWL1 is to express qualified cardinality, relational expressivity, and data type expressivity purpose, thus, OWL 2 is required (Grau et al., 2008).

2.1.10 *Various tools are available for Ontology developed:*

- OntoEdit (Sure et al., 2002) provides an additional feature along with ontology construction like inferencing and collaboration facilities.
- OilEd (Bechhofer et al., 2001) is OIL based ontology editor used to create and edit ontologies.
- WebODE (Corcho et al., 2002) provide an integrated platform for ontology engineering where many activities like creating, exchange and reasoning of ontologies occur.
- Ontolingua (Farquhar, 1997) allow creating and editing of ontologies in distributed collaborative environment.
- Ontosaurus (Corcho et al., 2003) is server based ontology development tool where the Ontosaurus ontology server uses Loom for its Knowledge representation.
- LinkFactory (Ceusters et al., 2001) manages the ontology building and also provide a solution by integrating information from an external relational database to their LinkBase® Ontology.
- Protégé (Noy and McGuinness, 2001) is a free and open source tool to build an ontology, developed by Mark Musen's group at Standford Medical Informatics.
- ALTOVA (Altova, 2012) is a commercial software for ontology development and integrated development environment (IDE).

2.1.11 *Ontology Query Language and Reasoners:*

- SPARQL query language is used to query data stored in an ontology (DuCharme, 2013).
- Fact++ (Tsarkov and Horrocks, 2006) is a description logic based reasoner.
- Pellet (Sirin et al., 2007) is an OWL DL reasoner written in Java language.
- HermiT (Shearer et al., 2008) is a fast OWL reasoner based on hyper- tableau calculus.
- CEDAR (Amir and Ait-kaci, 2014), old CEDAR version only provides taxonomy reasoner, but a new CEDAR version has the power to express DL-based relational roles.

**2.2 *Latent Semantic Indexing (LSI)***

LSI is based on the vector-retrieval method in which predefine relationship between terms is modeled (Deerwester et al., 1990). The advantage the usage of LSI algorithm provides is twofold: it allows us semantic querying and also respects the interrelatedness of the terms within a document.
- Many search engines depend on the syntactic search rather than semantic search for retrieving results to user's query terms. For example, a search engine like Google, Bing, Yahoo!, etc., uses user's query literal terms to retrieve results. This may sometimes lead to an irrelevant result as the one term may have several different meanings. The probability of failing to retrieve documents that don't contain user's query literal terms, but are meaningful in the intended context of the user's query, is higher. The LSI provides for the retrieval of the relevant document, even if it does not literally match the user's query terms in the particular document.
● Moreover, various models like Boolean, standard vector and probabilistic consider user literal terms as independent; this is a barrier to retrieve the relevant document. Terms in a document can't possibly be treated as independent due to interrelatedness. The term–term interrelation can actually help to improve retrieval.

LSI measure the similarity of context in which the word appears and creates a reduced dimension feature-space representation. It identifies associative relationships, useful in indexing and for the retrieval of information, using SVD, a mathematical technique closely related to eigenvector decomposition and factor analysis (Rosario, 2000). Information retrieval is done by lexical matching of the terms in a user's query with the terms in the documents. But LSI gives inappropriate results when a user's query is matched literally with the terms in a document, since different words may have similar meanings (synonym), and words may have multiple meanings. LSI overcomes these drawbacks of information retrieval through lexical matching by mathematically derived abstract indices rather than using individual words for retrieval. The primary inspiration for the present research is to build the "Semi-Automatic Topic Ontology" model (Fortuna et al., 2006) where LSI or K-means clustering is opted to discover keywords and further integrated with an interactive platform which recommends topics.

Information Retrieval (Berry et al., 1995; Young, 1994), Relevance Feedback, Information Filtering, TREC and Cross-Language Retrieval are a few of the many application areas of LSI.

However, LSI is not without its disadvantages:
● LSI retrieve inappropriate results if user's query having different words with similar meanings (synonym) or words with multiple meanings.
● LSI model is a distributed model it difficult to take into account latent dimension.
● For nonlinear dependencies, LSI is not the best solution as it is a linear model.
● A number of dimensions depend on the rank of the matrix.
● LSI works without human intervention as words are compared for every statement in the latent space.
● LSI provides efficiency searching for a user's question answer in the term by retrieving a document is easy as LSI use associate inverted index in vector area.

**2.3** *probabilistic Latent Semantic Indexing (pLSI)*

In spite of the fact that LSA has been successful in distinctive areas including programmed indexing (LSI), it has various deficiencies, mostly because of the admissible factual establishment. Hofmann in 1999 presents a novel way to deal with LSA and element examination called Probabilistic Latent Semantic Analysis (PLSA) (Hofmann, 1999) that has a strong factual establishment. Since it takes into account the probability standard and characterizes a fitting generative model of the information. This infers in specific that standard strategies from measurements can be sought inquiries like model fitting, model mix, also, multifaceted nature control.

However, pLSI is not without its disadvantages:
● pLSI is not able to classify new unseen documents.
● The Bag-of-Words assumption.
● A contingent independence of words and reports, which are just coupled through the inactive variable.

Advantages of pLSI over LSI:
- pLSI has a strong factual establishment since it is in light of the probability guideline and characterizes a fitting generative model of the information (Hofmann et al., 1999).
- Because of a strong factual establishment, it can be requested inquiries like model fitting, model mix, and multifaceted nature control.
- The element representation got by PLSA permits to manage polysemous words and to expressly recognize diverse implications and distinctive sorts of word utilization.

Comparison of LSI and pLSI:
- Retrieval of information through pLSI is very quick than the LSI.
- Pre-processing time for pLSI is less than LSI.
- Since pre-processing time and retrieval time of pLSI is better than LSI. PLSI is better than LSI but precision and topic, choosing is not so good in pLSI. Though information retrieval is faster with pLSI, the probability of an incorrect result is high, therefore precision, accuracy, and recall are a major drawback of the pLSI. These drawbacks are overcome by many other information retrieval methods like LDA and PAM.

**2.4** *Latent Dirichlet Allocation (LDA)*

The latent (hidden) term describes the hidden topic among words. Once the topic is identified, managing, organizing, and annotation of large online data archives of text is done. This process may also be done using probabilistic modeling. LDA is a subclass of the probabilistic modeling of topics which uses Bayesian probability (Blei et al., 2003). LDA is better than the pLSI in the following terms:
- LDA is based on the three-level hierarchical Bayesian model.
- The topic distribution in LDA is supposed to have a Dirichlet prior their topics.
- LSI is a dimensionality reduction technique. Synonyms and polysemy are basic linguistic notions that can be captured by the derived features of LSI. Probabilistic LSI (pLSI) model is improved form of LSI (use Bayesian method) but pLSI suffers from the problem such as over fitting. Future issues also arise in assigning probability to unseen document (outside of the training set) but LDA has the ability to handle the unseen document.
- The disadvantage of the pLSI is hard clustering:

A very large feature set is generated when individual words are considered as features. A spectral Algorithm for LDA, LDA model can represent words from several topics not just from a single topic. Unsupervised LDA is used as words are observed and the corresponding topic is hidden. Comparison of following topic modeling is shown in Table 1.

     LDA as a model that can be lengthened and used in many ways example: Variational Bayesian LDA (VB LDA), Online LDA, MapReduce LDA (Mr.LDA), Unsupervised LDA (LDA- Entity Resolution), Supervised LDA (Labeled-LDA and Multi-Grain LDA) etc.

We chose Mr.LDA due to its ability to fulfill following requirements:
- NELL problem as data on the internet continuously growing per second for this purpose we require unsupervised learning approach and scalable.

- Manually converting Text to ontology (Text-to-Onto) is time-consuming and error-prone process for the same we require an automatic process. Hence, we require an automatic process.
- Moreover other topic modeling uses Gibbs sampling whereas Mr.LDA uses Variational inference which can be easily implemented in distributed environment.
- Mr.LDA expected to reduce the require time as MapReduce programming framework work in parallel, distribute environment.

**Table 1**
Comparison of following Topic Modeling:

| Parameter | Latent Semantic Indexing (LSI) | Probabilistic Latent Semantic Indexing (pLSI) | Latent Dirichlet Allocation (LDA) |
|---|---|---|---|
| Introduce by author and year | Deerwester et al., 1990 | Hoffman, 1999 | Blei et al., 2003 |
| Latent topic/ Latent Variable | Yes | Yes. pLSI approach uses a latent variable model that represents the document as mixtures of topics. As aspect model associates co-occurrence of data with an unobserved class. | Yes |
| Application areas | Data Clustering, Information Retrieval, Information visualization and Document classification. | Natural Language Processing, Machine Learning, Bioinformatics, Information Retrieval, and Filtering | Collections of data (text corpus), Collaborative filtering, content-based image retrieval, and Bioinformatics, Document classification. |
| Bag-of-Words (BOW) assumption | Yes | Yes | Yes |
| Dimensionality Reduction | Yes (Vector Space Model) | Yes (Probabilisitc Model) | Yes (Probabilistic Model) (Crain et al., 2012). |
| Disadvantage | LSI can't contain multiple topics. Vector representation suffers from the curse of dimensionality, sparseness, also neglect semantic relationship between words. | • Suffer from overfitting issues.<br>• A previously unseen document can't be assigned probability using pLSI. Therefore, pLSI is not a well-defined model.<br>• As the number of training documents increases so does the number of parameters in pLSI. Also, there is no way to predict an unseen document.<br>• The pLSI model has a problem with inappropriate generative semantics. | |
| Advantage | | • The pLSI outperformed LSI in the vector space model framework. | • The unseen document can be analysis by unsupervised LDA. LDA models documents as a mixture of multiple topics.<br>• Modularity & Extensibility<br>• LDA was a semantically consistent topic model and thus it attracted interest from NLP and machine learning communities.<br>• Compared to the pLSI model, LDA possesses the capability of producing semantic by treating the topic mixture distribution as a k-parameter hidden random variable rather than a large set of individual parameters which are explicitly linked to the training set; thus, LDA overcomes the overfitting problem and the problem of generating the new document in pLSI. |
| Consider the geometry of the latent space | | Empirical distribution | Smooth distribution. |
| Approach uses various technique | LSI & SVD | | K-L divergence, Laplace approximation, Variational approximation & Markov Chain Monte Carlo (MCMC) (Jordan, 1999). |

## 3. Proposed Approach

### 3.1 *Semi-Automatic Terminology Ontology Learning Using LSI & SVD*

In the present research, two approaches are examined for semi-automatic ontology learning using LSI & SVD and Mr.LDA as shown in Fig. 1.

The documents are collected for constructing domain specific automatic ontologies. Domain specific ontology plays a vital role in information retrieval field. The process of ontology construction is a tedious task, so it is natural to search a way to make this process automatic. To create automatic domain specific ontology from text documents, the document needs to be read and processed. Here document is considered to be set of words. LSI is a vector space approach. LSI catches term–term statistical reference by replacing the document space with lower dimensional concept space. For this purpose it uses a technique called SVD (Maddi et al., 2001). After removing stop words (unimportant words like the, is, a), a count of the frequency of meaningful terms is collected. This is called preprocessing of text. Once the frequency count is calculated, in the next step, the weight of each semantically significant term is calculated and normalized. The sequential step is to extracting latent concepts from documents using LSI & SVD for Ontology Learning.

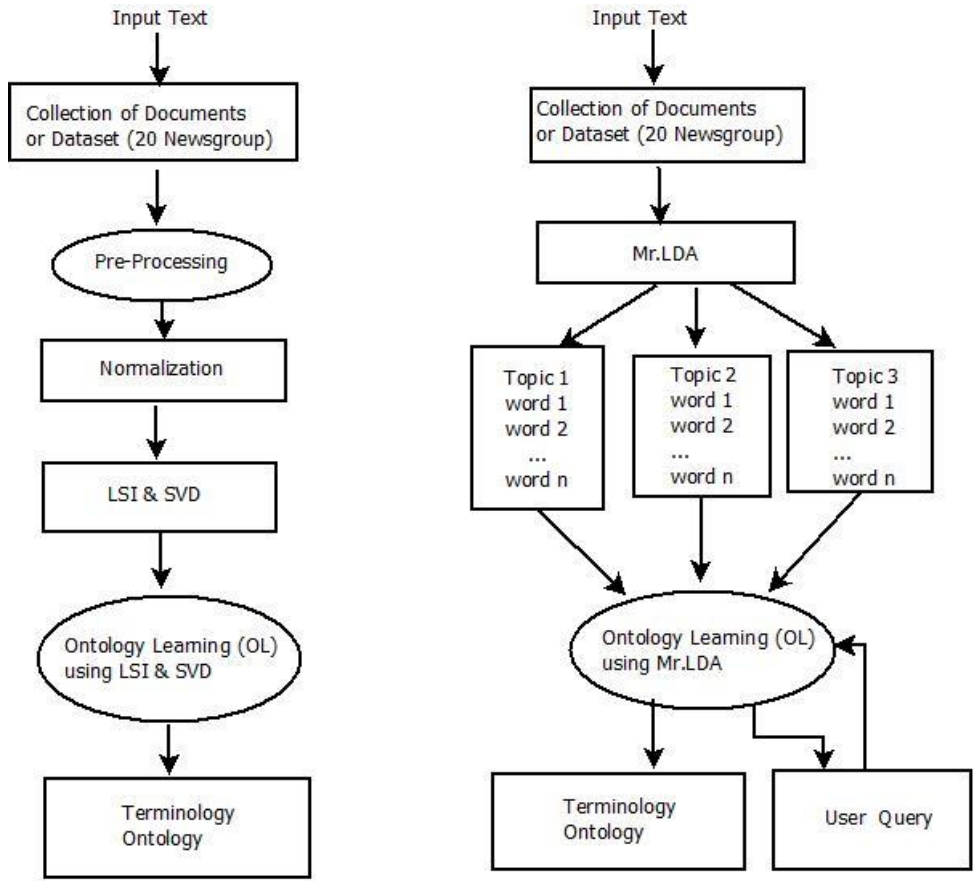

(a). LSI & SVD  (b). Mr.LDA
Fig. 1. Architecture of Ontology Construction Prototype

**3.2 *Semi-Automatic Terminology Ontology Learning Using Mr.LDA***

It is important to develop an ontology construction process which is based on the latent topic, to make sure the success of ontology creation and to come over the hindrances produce by old collections of history. The final ontology along with Sematic Web related standards are defined to represent the structure of the knowledge. Once the raw test is generated and corrected, the latent topics are extracted next. In this sequence, the LDA plays its role to find latent topics from text corpus that is newly generated in the previous sequence since LDA has an easier range calculation of topics than pLSI and LSI (Girolami and Kaban, 2003).

A statistical model like LDA is considered to determine whether the content of a document belongs to a particular topic. It is time-saving once the topic is known and becomes optional choice to further pursue in detail for particular fields. The LDA can be used to find cold or hot topics ($\Theta j$) and can be measured over time (year). The word assigned to topic help in analyzing topic dynamics per year. This quantitative information helps the novice researcher in either particular domain or even helpful for historical purposes for deep study over time. The rise and fall trend analysis of a particular topic can be measured for the purpose of identifying popularity of the topic in the research domain. The semantic of content can be analyzed and sum up by assigning a topic to it. The LDA can be used to collect documents starting from e-mail records, newsgroups to the entire internet (Griffiths and Steyvers, 2004).

The objective is to analyze words obtained from SNA (Social Network Analysis) or Gmail or distinguishing documents, such as emails and academic publication. Data mining models have been used in many SNA models. The computing tool Bash Reduce for word processing was used in parallel and generate connected words under every lurking topic to find conventional information on political news sending especially to local Columbus receptors. Higher precision is shown by LDA model than by TF-IDF model. SNA model put into action to make connections of some group by directing graph models or weighted edge graph models. For example, Wang and his co-worker (Wang et., 2010a) developed a probabilistic factor graph model to evaluate bibliographic network in the intellectual area. Ontology (directed graph models) or Fuzzy ontology (Weighted edge graph model) helps us to have a deep understanding of a topic in real world problem. Generally, e-mail is downloaded and developed to input data for the LDA model in a suitable format. LAD model is a Bayesian hierarchy topic model, creating topic words for every document with accurately reduce complexity. The LDA model describes the possibility that one document may have various topics while unigram model focuses on the single topic situation. Instead of calculating the probability of word frequency by continually multiplying in unigram model, LDA model maps a document of the N words to latent topics. The capability of LDA model can be used to automatically classify topics and words of e-mails which allow detection of fraud e-mails thus can be helpful for predictive analysis and security purpose. This process can be paralleled using MapReduce method along with LDA which reduce time and cost (Qiu and Stewart, 2014).

As the online data increases manifolds by the day, searching particular information under a certain domain becomes a difficult task. The topic label may help readers to find the highly impressive part of the text the corpus. Readers trying to find the most interesting part of a document are facilitated using topic labels. Algorithms that target to classify a large set of the corpus, having a thematic information have been developed by machine learning researchers. The LDA model determines the topic for a massive collection of unstructured documents. LDA is an unsupervised approach where the labeling of documents is not required. But for text analysis, the extra acquired information such as geographic location, title, author and others are incorporated as metadata. Validation must be provided to the unsupervised learning. LAD is used to retrieve words with the highest probability. The LDA probabilistic model can work on multiple topics, for example, various sections of the newspaper like Classified ads, Business/finance/economy, Automobiles, International news, Health/Medicine, Letters from the readers, Magazines, Politics (Blei, 2012). A text corpus suffers from the curse of dimensionality which can diminish using the LDA model. For dimensionality reduction, a topic model is used as an efficient technique. It can identify the semantic relationship between topic-document and word-topic. Automated ontology learning which is considered as an approach for gaining knowledge from the unstructured corpus. Being resistant to obsolescence, cost-effectiveness, and fast development is the advantage of an automatic building of ontologies. The terms "concept hierarchy" and "terminological ontology" can be interchanged in few texts. For finding information document modeling, classification purposes, and topic extraction are used in probabilistic topic models. The feasibility of using topic models in ontology learning was examined. In the research area of computer science, ontology conceptual represents a specific area of knowledge. Ontologies can be classified as terminological ontologies, prototype-based, and formal ontologies. The ontology learning can be further classified into six subtasks: relations, rules, concept hierarchies' synonyms, concepts and learning terms. This classification into subtask decreases the time and effort that generally in the ontology engineering process which uses human expertise (Wang et al., 2010b).

The Cosine of Topic Pointwise Mutual Information (CosTMI) (Lin et al., 2012) calculates the semantic closeness among topics and establishes these topics into the hierarchical model, hence forming the latest ontology. Manual ontology establishment has been a labor-intensive, time-consuming and costly work for developers. These drawbacks have been able to draw huge attention for semi-automatic ontology learning. Many of the known ontology learning methods concentrate on enriching or expanding prototype ontology with new things derived from the document. There are some methods of self-ontology learning, that have a few limits. An Ontology Learning Automatically (AOL) was assumed to learn concepts ontology on its own from provided documents. The LDA model was used which can create topics shown as multinomial spreading's over terms which act as the concepts of the ontology. To establish the connections among these topics, the present research explains CosTMI option to organize topics in stratified structure to create an ontology. The Method of AOL does not need prior information about a special field. Except for a single corpus of the document, it does not need ontology or any additional resource. The advantage of using AOL as it only requires a text corpus of various domains to create an ontology. LDA is used to find the concepts related to ontology from a corpus of the text document, which is given and then one after another utilize the model to learn the stratified concepts for creating the hierarchical model. A topic model is a creative model which is probabilistic and has been proved to be the accurate method for finding concepts without requiring any initial information. The relationships between concepts are identified as the useful approach to AOL also dynamic approach can be followed to learn ontology.

We analyze three algorithms for LDA:

● Variational Bayesian (VB) inference.

● Online Variational Bayesian inference

● Markov Chain Monte Carlo (MCMC)

On comparing the complications and the working of the above-mentioned approaches, it was found that Online VB inference is the fastest among the three, and online VB inference gives related results in comparison to VB inference and MCMC. The presence of large corpora necessitates the automatic inference of topics (Špeh et al., 2013). Online variational Bayes which is created for LDA is an algorithm that needs only a little extra line of code than the folk batch VB and is used for huge collections of documents. For LDA, Online VB approximates previous as well as the posterior ideas in a fraction of the expected time. The approach that has been used to obtain for LDA, an online version of the batch, VB is not complicated to use, for a large variety of hierarchical Bayesian models (Hoffman et al., 2010). LDA could not extract relevance among topics, however Correlation Topic Modeling (CTM) can be used for this purpose. LDA could not find relevance among topics as it only shows high probability words and corresponding relevance (Jing et al., 2012).

CTM solves the inability of LDA, as the latter is unable to design correlation topic. The correlation topic model created on initial LDA model which is a fragment of a family and has a mixed membership model for disintegrating data into various latent components. Restriction of a topic model like LDA is the incapability of directly modeling correlation among topics. This restriction for the LDA Model stems from a free assumption defined in the LDA distribution over the topical parts. Under LDA, the parts of the vectors that are proportional are not dependent, which leads to the ideal and the strong modeling assumption that the occupancy of one topic is not related to the occupancy of another. In fact, the capability to design correlation among the topics decreases some of the convenient computations that LDA affords. Particularly, the relation among the Dirichlet and multinomial allows straight approximate inference in LDA. When the Dirichlet is changing with a logistic-normal that conjugacy is lost. A standard simulation technique like Gibbs sampling are impossible, and Metropolis-Hasting is restricted due to the high dimension and scale of the data. Therefore, a quickly changeable method for making close inference possible with the CTM is created. CTM is a stratified model of the document collection. The words of each document from mixture model are models of CTM model. The component mixture is also used by corpus in the collection. The CTM allows each corpus to show various topics with varying proportions. It can catch the heterogeneity in contained text data that shows various latent patterns. The CTM created on LDA model. The advantage, in fact, is that it gives the document model which is more expressive. The LDA-based model will suppose things based on the lurking subjects that are suggested by the observations, but CTM will assume things related to extra topics that are related to conditionally probable topics. This (CTM) lets us embed newly realized, documents corpus into low dimensional lurking the mantic space which is shown by the model. A quick variational inference algorithm is used to almost accurate this posterior that lets us quickly analyze a large collection of documents under the difficult modeling assumption. Topic graphs: The capability of the CTM to design the relation between topics gives a superior fit of a corpus collection than LDA. The covariance of the logistic-normal model for the proportion of the topic can be used to anticipate the relationship between the topics. Specifically, the covariance matrix may be used to create a topic graph, where particular topics are shown by the node and surrounding nodes shows deeply connected topics. Recommendation of similar articles from a large number and volume, but the users of the journal would benefit from articles that are topically organized by using co-relational topic analysis.

In few new scholarly electronic fields, such as the ArXiv, the contributors give metadata along with manuscripts that explain and distribute the proposed work to aid in exploring the topics in the collection. In profuse text data sets, data that is metadata is not available. Acknowledge of structure takes place at the document level, where a latent vector of proportions (topic) is related to each document. The posterior spreading of the amounts may be used to relate document with topics that are latent. The log probability of stored info given a model assumed from the left data is calculated. The document collection that is a better model will allocate higher probability of the data that is stored. The average held out log probability (HOLP) for every model and average difference among a number of topics. CTM always gives a superior fit.

Another significant calculation of linked strength of LDA and the CTM is how better the model assumes the remaining words in a document after analyzing a part of it. When little numbers of words have been taken into consideration, there is not more uncertainty about the words that remained under the CTM in comparison to LDA. CTM uses relation among topics to differentiate that words in a linked topic can also be possible. In comparison, until a huge part of the document corpus has been thoroughly analyzed such that all of its topics are represented, LDA cannot guess the remaining words well. A stratified topic model of text corpus that changes the Dirichlet distribution of each document-topic portions with a logistic normal were created. CTM allows the model to track the correlation among occurrence of latent topics. The resultant correlation topic model uncovers interesting, detailed statistics for helping searching and browsing and gives superior performance. Use of logistic-normal may have the advantage in the various applications of Dirichlet-based group membership models, while more complex. One issue that was not thoroughly explored is model selection, which is choosing the number of topics for a collection. In another topic model, the non-parametric Bayesian method based on the process of Dirichlet is a natural tools suite because it gathers new topics as more text corpus are analyzed. Logistic normal, in fact, does not instantaneously provide a way to such extensions. Accepting the selection model issues in this type of setting is an important area for the research in the future (Blei and Lafferty, 2007). Following are application area of LDA:

- Association for Computational Linguistics (Ramage et al., 2009).
- Sentiment Analysis of product reviews (Lakkaraju et al., 2011).
- Probability Fuzzy ontology (Lau et al., 2014).
- Multilingual news analysis (Dubey et al., 2011).
- Correlation Topic Model (CTM), Logistic normal prior with full co-variance matrix (Blei and Lafferty, 2006a).
- Analysis using LDA: Trend over Time (Blei and Lafferty, 2006b).
- A LDA for unsupervised entity resolution (LDA-ER) - entity resolution uses the topical interest to determine if two author names correspond to the same author, LDA: topic mixtures independent of author names, ATM: Topic mixture distinct for different author names and identical for same author names (Bhattacharya and Getoor, 2006).
- Learning is hard, ploytime (uncountable) under restrictions (Anandkumar et al., 2014).
- The inference is hard (in general) for LDA (Sontag and Roy, 2011).
- No theoretical justification for collapsed versions, but empirical gains observed (Newman et al., 2009).
- Non-negative matrix factorization (NMF) - Computing an NMF-provably (Arora et al., 2012).
- The Wei & Croft use LDA on TREC dataset to efficiently improve ad-hoc retrieval (Wei and Croft, 2006).
- Labeled LDA is better than SVM, Labeled LDA: Force mapping between topics and labels assigned to documents. A subset of topics observed for each document: Constrain topic labels for a document to come from this fixed set.
- The Lukins et al., explore a technique for the developer to search bug in source code using LDA (Lukins et al., 2008).
- L-LDA (Labeled-LDA) which is an extension of LDA provides action in every video. This is done by L-LDA which is also a supervised topic model, to provide the explanation of visual words. Type-2 Fuzzy Set (T2FS) is used to define uncertain parameters in L-LDA. The capability of T2 L-LDA to represent three dimensions, it well suits for human action categorization problem (Cao and Liu, 2012).

Yeh and Yang provide corrections that are statistical and also a data of the text corpus are extracted according to the latent topic. This helps in producing a semantic-oriented and OWL-based ontology. As human beings are able to generate an efficient semantic hierarchy, manual construction of ontology has been a hard task. But with building connections and concepts, it is not easy for humans to create and maintain ontologies on the large scale. In the meantime, the structural quality of the knowledge in an ontology is very difficult to maintain as it is unable to have the way of consistent concept creation. Not a long time from today, the researches about the topic eradication of from text are becoming popular. The latent discovery of the topic is created to overcome the processing model of Bag-of-Words (BoW) in information gaining area (Yeh and Yang, 2008).

These text corpus of documents will produce a document-term matrix know in the retrieval of information domain which is basically a sparse matrix. Then the estimation of LDA begins and the topics which are latent created. Topics which are latent are clubbed into topics of higher-level in a hierarchical way. Because the latent topics contain semantics, so the process of clustering is seen as some type of semantic clustering. The cosine which is basically similar with hierarchical agglomerative clustering (HAC) is produced to create topics of high-level known as 'super topics'. Structure and ontology layer in this research: super topic –>latent topic layer –>Index layer (subject, organization, and person) and at last layer of page document. The topics from the structure of a tree, but the complete ontology need to be a tree structure since the connection between latent topics, index terms and page documents are both graph-based and hierarchical.

After the introduction of protégé software, the generation of ontology result to domain experts has been prepared and provided, for more optimization which includes maintenance and revision. In future, the improvement on proposed algorithm of ontology processing to create ontology with superior semantic quality will be concentrated. For example, concept clustering is being looked into, in place of similarity of term vector in topic clustering step by step using similarity propagation of term relationship.

The LDA model shows more accurate results for question answering then our traditional model using Ontology Automatic question answering among on various platforms like (Quora/LinkedIn/ResearchGate) classify questionable on the bases of tag suggested by users. If the database is huge then ruled corpora for reasoning. This kind of platform immensely benefited students to search related information from a large information resource. When working with large corpora of documents it is difficult to retrieve accurate without document model like LDA. As LDA determining specific topics to which document is related this information help student to understand to proceed future or not in the quick span of time. Most of the information retrieval technique usually relies on word matching it mean only syntactic analysis, not semantic analysis. To do semantic analysis of large corpora of documents is a difficult task, for this purpose numerous approach, are available for feature selection technique. As some model like LDA uses the search latent feature of the larger document. These features (latent feature) can be used to make ontology, once ontology is ready it gives semantic meaning to information. Ontology-based scheme for the annotation of a document is used for semantic retrieval. Using latent feature author constructs a semi-automatic topic ontology (fuzzy ontology) to enhance accurate semantic search rather than syntactic search. Information retrieval and ontology both fields are kept on combining and emerging field for an accurate semantic search. Integrating ontologies help users to retrieve a more accurate document and personalized (Hu et al., 2014).

For user's search keyword (user's query key terms) if there is synonym present, then the user's query can be expanded and one the basis of the semantic similarity document is retrieved. For this purpose, WordNet is used to retrieve synonym terms and find the ontology. A document is presented as high dimensional when each word is seen as a feature. It's often necessary to use dimension reduction technique as a presentation of the document has been higher dimensioned. Here, Fuzzy Co-Clustering and Single Term Fuzzier (FCC_STF) was used to curse the dimensional reduction (Rani et al., 2012). A latent representation space of document semantics is made of a linear transform of word count using LSI. PLSI uses LSI with a general expectation maximization algorithm using this model deals with domain specific synonyms and polysemous words. The LDA topic model not only efficiently retrieves an accurate document but also capable of retrieving text segment for question answering this capability can be applied to any domain like Agriculture, Medicine, and Gene Ontology etc.

Mr.LDA model performs parallel processing for each segment for this purpose share memory is required to keep consistency. Mr.LDA more effectively handles online batches which make it easier to learn largely collected document corpus (Zhai et al., 2011). Variational inference generally needs plenty of iterations to assemble, while Gibbs sampling requires thousands. This is because Gibbs sampling needs more synchronizations to finish inference, adding to the intricacy of the communication overhead and the implementation. Whereas varying inference needs synchronization per iteration (many times for a specific document) in a sampling implementation, inference needs synchronization after each word in each iteration. Anonymity: By definition, Monte Carlo algorithm depends on anonymity. In fact, MapReduce implementations supposedly take each computational step will be exact, despite the fact when or where it is run. This leads to greater tolerance of fault, running many sets of computation subcomponents either one is taking too long or not true. The variational method reduces the high dimensional problem and gives specific advantages when latent variable pairs are non-conjugate. Mr.LDA's modular nature makes the design flexible. Mr.LDA may be enhanced to more effectively handle online batches which make it easier to learn large document collections (Zhai et al., 2012).

## 4. Experimental Result

### 4.1 *Experimental results using LSI & SVD*

The description of the LSI & SVD topic modeling consists of following steps:
Step 1: Pre-processing
From the dataset (20 Newsgroup) or input text documents, meaningful terms are extracted and frequencies are calculated this process is called Pre-processing. Results are shown in Fig. 2. Pre-processing transforms the input text document into a term-document matrix in LSI database which enables meaningful statistical analysis. Terms and document matrix are stored in LSI database which involves SVD of a normalized weight matrix.

| | |
|---|---:|
| word 45 = imported | frequency = 1 |
| word 46 = Holland | frequency = 1 |
| word 47 = kick | frequency = 1 |
| word 48 = local | frequency = 1 |
| word 49 = boy | frequency = 1 |
| word 50 = james | frequency = 1 |
| word 51 = ii | frequency = 1 |
| word 52 = provided | frequency = 1 |
| word 53 = basis | frequency = 1 |
| word 54 = denoument | frequency = 1 |
| word 55 = introduced | frequency = 1 |
| word 56 = errol | frequency = 1 |
| word 57 = world | frequency = 1 |
| word 58 = love | frequency = 1 |
| word 59 = interest | frequency = 1 |
| word 60 = Olivia | frequency = 1 |
| word 61 = de | frequency = 1 |
| word 62 = Havilland | frequency = 1 |
| word 63 = Flynn | frequency = 2 |
| word 64 = film | frequency = 2 |
| word 65 = exercise | frequency = 2 |
| word 66 = movie | frequency = 2 |
| word 67 = buffs | frequency = 2 |
| word 68 = films | frequency = 2 |
| word 69 = expressed | frequency = 1 |
| word 70 = Theodore | frequency = 2 |
| word 71 = major | frequency = 1 |
| word 72 = university | frequency = 1 |
| word 73 = remus | frequency = 2 |
| word 74 = Rutgers | frequency = 5 |
| word 75 = kaldis | frequency = 5 |
| word 76 = hold | frequency = 1 |
| word 77 = views | frequency = 2 |

Fig. 2. Pre-processing

Step 2: Create and display Binary matrix
Fig. 3 depicts the results of Create and display Binary matrix.

```
Row689 ====   0  0  0  0  0  0  0  0  1    word = replace
Row690 ====   0  0  0  0  0  0  0  0  1    word = queen
Row691 ====   0  0  0  0  0  0  0  0  1    word = dimension
Row692 ====   0  0  0  0  0  0  0  0  1    word = restrictive
Row693 ====   0  0  0  0  0  0  0  0  1    word = privileges
Row694 ====   0  0  0  0  0  0  0  0  1    word = theoretical
Row695 ====   0  0  0  0  0  0  0  0  1    word = iii
Row696 ====   0  0  0  0  0  0  0  0  1    word = imported
Row697 ====   0  0  0  0  0  0  0  0  1    word = Holland
Row698 ====   0  0  0  0  0  0  0  0  1    word = kick
Row699 ====   0  0  0  0  0  0  0  0  1    word = local
Row700 ====   0  0  0  0  0  0  0  0  1    word = boy
Row701 ====   0  0  0  0  0  0  0  0  1    word = james
Row702 ====   0  0  0  0  0  0  0  0  1    word = ii
Row703 ====   0  0  0  0  0  0  0  0  1    word = denoument
Row704 ====   0  0  0  0  0  0  0  0  1    word = introduced
Row705 ====   0  0  0  0  0  0  0  0  1    word = errol
Row706 ====   0  0  0  0  0  0  0  0  1    word = love
```

```
Row707 ====  0 0 0 0 0 0 0 0 1     word = Olivia
Row708 ====  0 0 0 0 0 0 0 0 1     word = Havilland
Row709 ====  0 0 0 0 0 0 0 0 2     word = Flynn
Row710 ====  0 0 0 0 0 0 0 0 2     word = film
Row711 ====  0 0 0 0 0 0 0 0 2     word = exercise
Row712 ====  0 0 0 0 0 0 0 0 2     word = movie
Row713 ====  0 0 0 0 0 0 0 0 2     word = buffs
Row714 ====  0 0 0 0 0 0 0 0 2     word = films
Row715 ====  0 0 0 0 0 0 0 0 1     word = expressed
Row716 ====  0 0 0 0 0 0 0 0 2     word = Theodore
Row717 ====  0 0 0 0 0 0 0 0 1     word = major
Row718 ====  0 0 0 0 0 0 0 0 2     word = remus
Row719 ====  0 0 0 0 0 0 0 0 5     word = Rutgers
Row720 ====  0 0 0 0 0 0 0 0 5     word = kaldis
Row721 ====  0 0 0 0 0 0 0 0 2     word = views
```
Fig. 3. Create and display Binary matrix

Step 3: Create and display Terms Weight Matrix

The frequencies of words in a document obtained during pre-processing. Then the weight is calculated. Term Weight is calculated by the following formula: Where $W_{i,k}$ is the weight of the ith term in kth document and $n_k$ is the total number of terms in that document. The obtained matrix represents terms and documents as rows and columns respectively and known as term-document matrix. Fig. 4 depicts the results of Create and display Terms Weight Matrix. For creating term-document matrix various documents and terms occurred are considered in input text files (Dataset).

```
Row689 ====  0.000 0.000 0.000 0.000 0.000 0.000 0.000 0.000 0.010     word = replace
Row690 ====  0.000 0.000 0.000 0.000 0.000 0.000 0.000 0.000 0.010     word = queen
Row691 ====  0.000 0.000 0.000 0.000 0.000 0.000 0.000 0.000 0.010     word = dimension
Row692 ====  0.000 0.000 0.000 0.000 0.000 0.000 0.000 0.000 0.010     word = restictive
Row693 ====  0.000 0.000 0.000 0.000 0.000 0.000 0.000 0.000 0.010     word = privileges
Row694 ====  0.000 0.000 0.000 0.000 0.000 0.000 0.000 0.000 0.010     word = theoretical
Row695 ====  0.000 0.000 0.000 0.000 0.000 0.000 0.000 0.000 0.010     word = iii
Row696 ====  0.000 0.000 0.000 0.000 0.000 0.000 0.000 0.000 0.010     word = imported
Row697 ====  0.000 0.000 0.000 0.000 0.000 0.000 0.000 0.000 0.010     word = holland
Row698 ====  0.000 0.000 0.000 0.000 0.000 0.000 0.000 0.000 0.010     word = kick
Row699 ====  0.000 0.000 0.000 0.000 0.000 0.000 0.000 0.000 0.010     word = local
Row700 ====  0.000 0.000 0.000 0.000 0.000 0.000 0.000 0.000 0.010     word = boy
Row701 ====  0.000 0.000 0.000 0.000 0.000 0.000 0.000 0.000 0.010     word = james
Row702 ====  0.000 0.000 0.000 0.000 0.000 0.000 0.000 0.000 0.010     word = ii
Row703 ====  0.000 0.000 0.000 0.000 0.000 0.000 0.000 0.000 0.010     word = denoument
Row704 ====  0.000 0.000 0.000 0.000 0.000 0.000 0.000 0.000 0.010     word = introduced
Row705 ====  0.000 0.000 0.000 0.000 0.000 0.000 0.000 0.000 0.010     word = errol
Row706 ====  0.000 0.000 0.000 0.000 0.000 0.000 0.000 0.000 0.010     word = love
Row707 ====  0.000 0.000 0.000 0.000 0.000 0.000 0.000 0.000 0.010     word = olivia
Row708 ====  0.000 0.000 0.000 0.000 0.000 0.000 0.000 0.000 0.010     word = havilland
Row709 ====  0.000 0.000 0.000 0.000 0.000 0.000 0.000 0.000 0.020     word = flynn
Row710 ====  0.000 0.000 0.000 0.000 0.000 0.000 0.000 0.000 0.020     word = film
Row711 ====  0.000 0.000 0.000 0.000 0.000 0.000 0.000 0.000 0.020     word = exercise
Row712 ====  0.000 0.000 0.000 0.000 0.000 0.000 0.000 0.000 0.020     word = movie
Row713 ====  0.000 0.000 0.000 0.000 0.000 0.000 0.000 0.000 0.020     word = buffs
Row714 ====  0.000 0.000 0.000 0.000 0.000 0.000 0.000 0.000 0.020     word = films
Row715 ====  0.000 0.000 0.000 0.000 0.000 0.000 0.000 0.000 0.010     word = expressed
```

```
Row716 ====  0.000 0.000 0.000 0.000 0.000 0.000 0.000 0.000 0.020    word = theodore
Row717 ====  0.000 0.000 0.000 0.000 0.000 0.000 0.000 0.000 0.010    word = major
Row718 ====  0.000 0.000 0.000 0.000 0.000 0.000 0.000 0.000 0.020    word = remus
Row719 ====  0.000 0.000 0.000 0.000 0.000 0.000 0.000 0.000 0.049    word = rutgers
Row720 ====  0.000 0.000 0.000 0.000 0.000 0.000 0.000 0.000 0.049    word = kaldis
Row721 ====  0.000 0.000 0.000 0.000 0.000 0.000 0.000 0.000 0.020    word = views
```
Fig. 4. Create and display Terms Weight Matrix

Step 4: Create and display Normalized Weight Matrix
For concept extraction normalization of the term weight matrix is required for every document present in dataset. The normalized term weights, collectively form the matrix W, where $W_{i,k}$ = NormalizedWeight$_{i,k}$ (using Eq.(2)) and Fig.5 represents the results.

$$W_{i,k} = \frac{frequency_{i,k}}{(\sum_{j=1}^{nk} frequency_{j,k})} \qquad (1)$$

$$\text{NormalizedWeight}_{i,k} = \frac{W_{i,k}}{\sqrt{\sum_{j=1}^{nk} W_{i,k}^2}} \qquad (2)$$

```
Row689 ====  0.000 0.000 0.000 0.000 0.000 0.000 0.000 0.000 0.075    word = replace
Row690 ====  0.000 0.000 0.000 0.000 0.000 0.000 0.000 0.000 0.075    word = queen
Row691 ====  0.000 0.000 0.000 0.000 0.000 0.000 0.000 0.000 0.075    word = dimension
Row692 ====  0.000 0.000 0.000 0.000 0.000 0.000 0.000 0.000 0.075    word = restictive
Row693 ====  0.000 0.000 0.000 0.000 0.000 0.000 0.000 0.000 0.075    word = privileges
Row694 ====  0.000 0.000 0.000 0.000 0.000 0.000 0.000 0.000 0.075    word = theoretical
Row695 ====  0.000 0.000 0.000 0.000 0.000 0.000 0.000 0.000 0.075    word = iii
Row696 ====  0.000 0.000 0.000 0.000 0.000 0.000 0.000 0.000 0.075    word = imported
Row697 ====  0.000 0.000 0.000 0.000 0.000 0.000 0.000 0.000 0.075    word = holland
Row698 ====  0.000 0.000 0.000 0.000 0.000 0.000 0.000 0.000 0.075    word = kick
Row699 ====  0.000 0.000 0.000 0.000 0.000 0.000 0.000 0.000 0.075    word = local
Row700 ====  0.000 0.000 0.000 0.000 0.000 0.000 0.000 0.000 0.075    word = boy
Row701 ====  0.000 0.000 0.000 0.000 0.000 0.000 0.000 0.000 0.075    word = james
Row702 ====  0.000 0.000 0.000 0.000 0.000 0.000 0.000 0.000 0.075    word = ii
Row703 ====  0.000 0.000 0.000 0.000 0.000 0.000 0.000 0.000 0.075    word = denoument
Row704 ====  0.000 0.000 0.000 0.000 0.000 0.000 0.000 0.000 0.075    word = introduced
Row705 ====  0.000 0.000 0.000 0.000 0.000 0.000 0.000 0.000 0.075    word = errol
Row706 ====  0.000 0.000 0.000 0.000 0.000 0.000 0.000 0.000 0.075    word = love
Row707 ====  0.000 0.000 0.000 0.000 0.000 0.000 0.000 0.000 0.075    word = olivia
Row708 ====  0.000 0.000 0.000 0.000 0.000 0.000 0.000 0.000 0.075    word = havilland
Row709 ====  0.000 0.000 0.000 0.000 0.000 0.000 0.000 0.000 0.151    word = flynn
Row710 ====  0.000 0.000 0.000 0.000 0.000 0.000 0.000 0.000 0.151    word = film
Row711 ====  0.000 0.000 0.000 0.000 0.000 0.000 0.000 0.000 0.151    word = exercise
Row712 ====  0.000 0.000 0.000 0.000 0.000 0.000 0.000 0.000 0.151    word = movie
Row713 ====  0.000 0.000 0.000 0.000 0.000 0.000 0.000 0.000 0.151    word = buffs
```

```
Row714 ====  0.000 0.000 0.000 0.000 0.000 0.000 0.000 0.000 0.151    word = films
Row715 ====  0.000 0.000 0.000 0.000 0.000 0.000 0.000 0.000 0.075    word = expressed
Row716 ====  0.000 0.000 0.000 0.000 0.000 0.000 0.000 0.000 0.151    word = theodore
Row717 ====  0.000 0.000 0.000 0.000 0.000 0.000 0.000 0.000 0.075    word = major
Row718 ====  0.000 0.000 0.000 0.000 0.000 0.000 0.000 0.000 0.151    word = remus
Row719 ====  0.000 0.000 0.000 0.000 0.000 0.000 0.000 0.000 0.377    word = rutgers
Row720 ====  0.000 0.000 0.000 0.000 0.000 0.000 0.000 0.000 0.377    word = kaldis
Row721 ====  0.000 0.000 0.000 0.000 0.000 0.000 0.000 0.000 0.151    word = views
```
Fig. 5. Create and display Normalized Weight Matrix

Step 5: Display Singular Value Decomposition for Normalized Weight Matrix
LSI is a method to express each term and document as a vector with the elements corresponding to concept. We obtain concepts using LSI method which involves decomposition of Normalized Weight Matrix (W) using Singular Value Decomposition as show in Fig. 6. LSI approach is helpful in links terms into the semantic structure without syntax or human intervention.

```
 0.039   0.016  -0.063  -0.056   0.000  -0.077   0.059   0.073  -0.008
 0.039   0.016  -0.063  -0.056   0.000  -0.077   0.059   0.073  -0.008
 0.039   0.016  -0.063  -0.056   0.000  -0.077   0.059   0.073  -0.008
 0.039   0.016  -0.063  -0.056   0.000  -0.077   0.059   0.073  -0.008
 0.019   0.008  -0.031  -0.028   0.000  -0.038   0.029   0.037  -0.004
 0.039   0.016  -0.063  -0.056   0.000  -0.077   0.059   0.073  -0.008
 0.019   0.008  -0.031  -0.028   0.000  -0.038   0.029   0.037  -0.004
 0.039   0.016  -0.063  -0.056   0.000  -0.077   0.059   0.073  -0.008
 0.097   0.040  -0.156  -0.140   0.001  -0.192   0.147   0.183  -0.019
 0.097   0.040  -0.156  -0.140   0.001  -0.192   0.147   0.183  -0.019
 0.039   0.016  -0.063  -0.056   0.000  -0.077   0.059   0.073  -0.008
Matrix S

 1.279   0.000   0.000   0.000   0.000   0.000   0.000   0.000   0.000
 0.000   1.072   0.000   0.000   0.000   0.000   0.000   0.000   0.000
 0.000   0.000   1.003   0.000   0.000   0.000   0.000   0.000   0.000
 0.000   0.000   0.000   0.987   0.000   0.000   0.000   0.000   0.000
 0.000   0.000   0.000   0.000   0.974   0.000   0.000   0.000   0.000
 0.000   0.000   0.000   0.000   0.000   0.960   0.000   0.000   0.000
 0.000   0.000   0.000   0.000   0.000   0.000   0.944   0.000   0.000
 0.000   0.000   0.000   0.000   0.000   0.000   0.000   0.909   0.000
 0.000   0.000   0.000   0.000   0.000   0.000   0.000   0.000   0.804
Matrix V

 0.241   0.124   0.489   0.032   0.751  -0.286  -0.168   0.100   0.051
 0.333   0.339  -0.444  -0.093   0.101  -0.098  -0.360  -0.645  -0.046
 0.346   0.237   0.160   0.269   0.039   0.250   0.757  -0.299  -0.016
 0.298   0.411  -0.188   0.274   0.006   0.515  -0.280   0.536   0.043
 0.467  -0.507   0.054   0.065  -0.053   0.089  -0.121   0.023  -0.702
 0.465  -0.518  -0.036  -0.020  -0.061   0.092  -0.064  -0.042   0.704
 0.166   0.215   0.428  -0.771  -0.228   0.303  -0.061  -0.031  -0.018
 0.232   0.246   0.377   0.329  -0.605  -0.483  -0.181   0.018   0.052
 0.328   0.113  -0.416  -0.368   0.003  -0.490   0.368   0.440  -0.042
```
Fig. 6. Display Singular Value Decomposition for Normalized Weight Matrix

Step 6: Find Concepts

LSI is the method used for extracting latent concepts from documents (20 Newsgroup) as shown in Fig. 7. LSI method represents the each document as concepts not as a set of terms, as concepts are statistically independent whereas terms are not.

| | | | | | | | | | | | | | | | | | | |
|---|---|---|---|---|---|---|---|---|---|---|---|---|---|---|---|---|---|---|
| 0 | 0.0002 | 0 | -0.0001 | 0 | -0.0000 | 0 | 0.0001 | 0 | -0.0022 | 0 | 0.0022 | 0 | -0.0001 | 0 | 0.0002 | 0 | 0.0752 | replace |
| 0 | 0.0002 | 0 | -0.0001 | 0 | -0.0000 | 0 | 0.0001 | 0 | -0.0022 | 0 | 0.0022 | 0 | -0.0001 | 0 | 0.0002 | 0 | 0.0752 | queen |
| 0 | 0.0002 | 0 | -0.0001 | 0 | -0.0000 | 0 | 0.0001 | 0 | -0.0022 | 0 | 0.0022 | 0 | -0.0001 | 0 | 0.0002 | 0 | 0.0752 | dimension |
| 0 | 0.0002 | 0 | -0.0001 | 0 | -0.0000 | 0 | 0.0001 | 0 | -0.0022 | 0 | 0.0022 | 0 | -0.0001 | 0 | 0.0002 | 0 | 0.0752 | restictive |
| 0 | 0.0002 | 0 | -0.0001 | 0 | -0.0000 | 0 | 0.0001 | 0 | -0.0022 | 0 | 0.0022 | 0 | -0.0001 | 0 | 0.0002 | 0 | 0.0752 | privileges |
| 0 | 0.0002 | 0 | -0.0001 | 0 | -0.0000 | 0 | 0.0001 | 0 | -0.0022 | 0 | 0.0022 | 0 | -0.0001 | 0 | 0.0002 | 0 | 0.0752 | theoretical |
| 0 | 0.0002 | 0 | -0.0001 | 0 | -0.0000 | 0 | 0.0001 | 0 | -0.0022 | 0 | 0.0022 | 0 | -0.0001 | 0 | 0.0002 | 0 | 0.0752 | iii |
| 0 | 0.0002 | 0 | -0.0001 | 0 | -0.0000 | 0 | 0.0001 | 0 | -0.0022 | 0 | 0.0022 | 0 | -0.0001 | 0 | 0.0002 | 0 | 0.0752 | imported |
| 0 | 0.0002 | 0 | -0.0001 | 0 | -0.0000 | 0 | 0.0001 | 0 | -0.0022 | 0 | 0.0022 | 0 | -0.0001 | 0 | 0.0002 | 0 | 0.0752 | holland |
| 0 | 0.0002 | 0 | -0.0001 | 0 | -0.0000 | 0 | 0.0001 | 0 | -0.0022 | 0 | 0.0022 | 0 | -0.0001 | 0 | 0.0002 | 0 | 0.0752 | kick |
| 0 | 0.0002 | 0 | -0.0001 | 0 | -0.0000 | 0 | 0.0001 | 0 | -0.0022 | 0 | 0.0022 | 0 | -0.0001 | 0 | 0.0002 | 0 | 0.0752 | local |
| 0 | 0.0002 | 0 | -0.0001 | 0 | -0.0000 | 0 | 0.0001 | 0 | -0.0022 | 0 | 0.0022 | 0 | -0.0001 | 0 | 0.0002 | 0 | 0.0752 | boy |
| 0 | 0.0002 | 0 | -0.0001 | 0 | -0.0000 | 0 | 0.0001 | 0 | -0.0022 | 0 | 0.0022 | 0 | -0.0001 | 0 | 0.0002 | 0 | 0.0752 | james |
| 0 | 0.0002 | 0 | -0.0001 | 0 | -0.0000 | 0 | 0.0001 | 0 | -0.0022 | 0 | 0.0022 | 0 | -0.0001 | 0 | 0.0002 | 0 | 0.0752 | ii |
| 0 | 0.0002 | 0 | -0.0001 | 0 | -0.0000 | 0 | 0.0001 | 0 | -0.0022 | 0 | 0.0022 | 0 | -0.0001 | 0 | 0.0002 | 0 | 0.0752 | denoument |
| 0 | 0.0002 | 0 | -0.0001 | 0 | -0.0000 | 0 | 0.0001 | 0 | -0.0022 | 0 | 0.0022 | 0 | -0.0001 | 0 | 0.0002 | 0 | 0.0752 | introduced |
| 0 | 0.0002 | 0 | -0.0001 | 0 | -0.0000 | 0 | 0.0001 | 0 | -0.0022 | 0 | 0.0022 | 0 | -0.0001 | 0 | 0.0002 | 0 | 0.0752 | errol |
| 0 | 0.0002 | 0 | -0.0001 | 0 | -0.0000 | 0 | 0.0001 | 0 | -0.0022 | 0 | 0.0022 | 0 | -0.0001 | 0 | 0.0002 | 0 | 0.0752 | love |
| 0 | 0.0002 | 0 | -0.0001 | 0 | -0.0000 | 0 | 0.0001 | 0 | -0.0022 | 0 | 0.0022 | 0 | -0.0001 | 0 | 0.0002 | 0 | 0.0752 | olivia |
| 0 | 0.0002 | 0 | -0.0001 | 0 | -0.0000 | 0 | 0.0001 | 0 | -0.0022 | 0 | 0.0022 | 0 | -0.0001 | 0 | 0.0002 | 0 | 0.0752 | havilland |
| 0 | 0.0003 | 0 | -0.0003 | 0 | -0.0001 | 0 | 0.0003 | 0 | -0.0044 | 0 | 0.0044 | 0 | -0.0001 | 0 | 0.0003 | 0 | 0.1505 | flynn |
| 0 | 0.0003 | 0 | -0.0003 | 0 | -0.0001 | 0 | 0.0003 | 0 | -0.0044 | 0 | 0.0044 | 0 | -0.0001 | 0 | 0.0003 | 0 | 0.1505 | film |
| 0 | 0.0003 | 0 | -0.0003 | 0 | -0.0001 | 0 | 0.0003 | 0 | -0.0044 | 0 | 0.0044 | 0 | -0.0001 | 0 | 0.0003 | 0 | 0.1505 | exercise |
| 0 | 0.0003 | 0 | -0.0003 | 0 | -0.0001 | 0 | 0.0003 | 0 | -0.0044 | 0 | 0.0044 | 0 | -0.0001 | 0 | 0.0003 | 0 | 0.1505 | movie |
| 0 | 0.0003 | 0 | -0.0003 | 0 | -0.0001 | 0 | 0.0003 | 0 | -0.0044 | 0 | 0.0044 | 0 | -0.0001 | 0 | 0.0003 | 0 | 0.1505 | buffs |
| 0 | 0.0003 | 0 | -0.0003 | 0 | -0.0001 | 0 | 0.0003 | 0 | -0.0044 | 0 | 0.0044 | 0 | -0.0001 | 0 | 0.0003 | 0 | 0.1505 | films |
| 0 | 0.0002 | 0 | -0.0001 | 0 | -0.0000 | 0 | 0.0001 | 0 | -0.0022 | 0 | 0.0022 | 0 | -0.0001 | 0 | 0.0002 | 0 | 0.0752 | expressed |
| 0 | 0.0003 | 0 | -0.0003 | 0 | -0.0001 | 0 | 0.0003 | 0 | -0.0044 | 0 | 0.0044 | 0 | -0.0001 | 0 | 0.0003 | 0 | 0.1505 | theodore |
| 0 | 0.0002 | 0 | -0.0001 | 0 | -0.0000 | 0 | 0.0001 | 0 | -0.0022 | 0 | 0.0022 | 0 | -0.0001 | 0 | 0.0002 | 0 | 0.0752 | major |
| 0 | 0.0003 | 0 | -0.0003 | 0 | -0.0001 | 0 | 0.0003 | 0 | -0.0044 | 0 | 0.0044 | 0 | -0.0001 | 0 | 0.0003 | 0 | 0.1505 | remus |
| 0 | 0.0008 | 0 | -0.0007 | 0 | -0.0002 | 0 | 0.0007 | 0 | -0.0110 | 0 | 0.0110 | 0 | -0.0003 | 0 | 0.0008 | 1 | 0.3762 | rutgers |
| 0 | 0.0008 | 0 | -0.0007 | 0 | -0.0002 | 0 | 0.0007 | 0 | -0.0110 | 0 | 0.0110 | 0 | -0.0003 | 0 | 0.0008 | 1 | 0.3762 | kaldis |
| 0 | 0.0003 | 0 | -0.0003 | 0 | -0.0001 | 0 | 0.0003 | 0 | -0.0044 | 0 | 0.0044 | 0 | -0.0001 | 0 | 0.0003 | 0 | 0.1505 | views |

Fig. 7. Find Concepts

Step 7: Display Concepts and related words

A concept node is used to represent concept name generated automatically as a most frequent hyphenated set of terms and related terms to particular concepts. Fig. 8 depicts the result of concepts and related words.

word1 = people
word2 = crimes
word3 = federal
word4 = classes

word5 = wdstarr
word6 = mit
word7 = equality
word8 = laws
word9 = law
CONCEPT NAME = -federal-classes-law-laws-oppose
word0 = people
word1 = based
word2 = protection
word3 = equal
word4 = crimes
word5 = class
word6 = federal
word7 = classes
word8 = wdstarr
word9 = mit
word10 = equality
word11 = laws
word12 = law
CONCEPT NAME = -mail-merrimack-address-hill-organization
word0 = Merrimack
word1 = address
word2 = mail
CONCEPT NAME = -energy-population-radford-toronto-technet
word0 = energy
CONCEPT NAME = -rutgers-kaldis-ferigner-mcgill-ca
word0 = ferigner
word1 = Rutgers
word2 = kaldis

Fig. 8. Display Concepts and their related words

Step 8: Display Terminological Ontology Learning
    ● Constructing document ontology

The domain ontology is constructed using concept node and term nodes from term matrix (U) and document matrix (V), which is obtained from SVD.
    ● Graph construction

Terminological Ontology is a bipartite graph use to represent knowledge between Concept and terms as shown in Fig. 9.

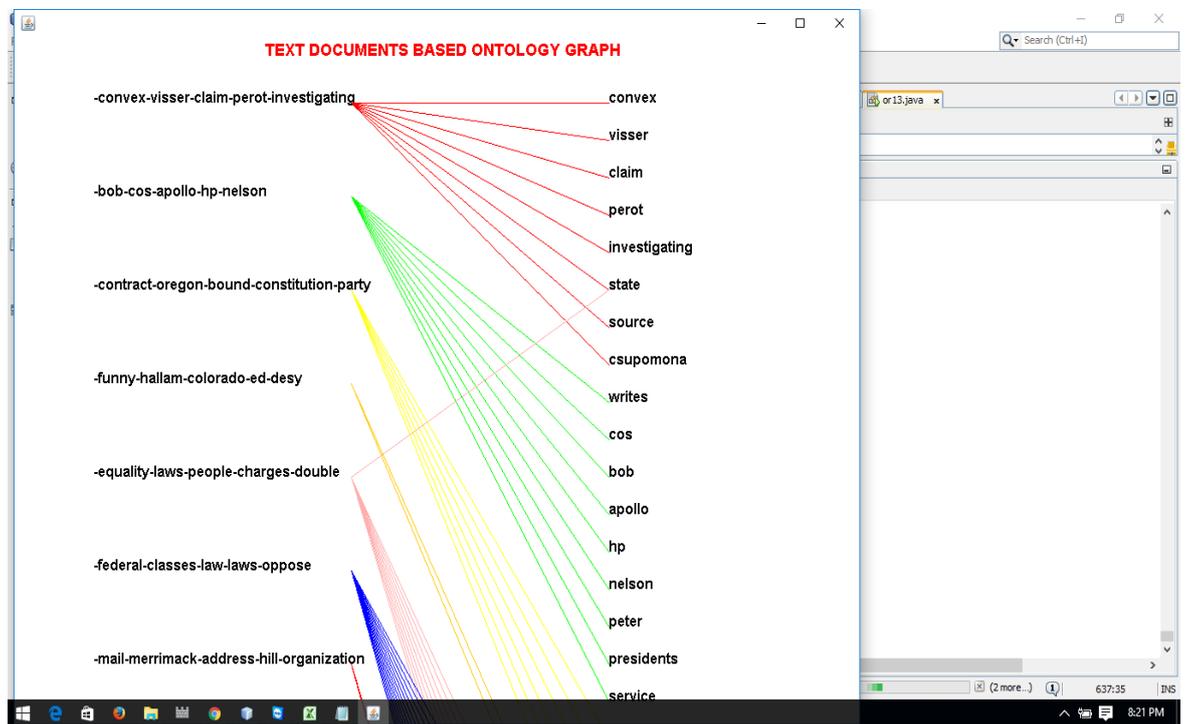

Fig. 9. Display Graph Data

Step 9: Once a new document is added to dataset then repeat above step (1-8) to learn terminological ontology.

### 4.2 Experimental results using Mr.LDA

#### 4.2.1 Tasks to perform for Mr.LDA

To accomplish this goal, the following tasks are to be performed:
Step 1: Tokenizing and pruning the text
The text used as an input is processed and tokenized, resulting in a set of nouns, adjectives, and verbs present in the input. Further, all the stop words like articles are removed.
Step 2: Topic Modeling
The Tokenized and pruned text are then subjected to the topic modeling algorithm called Mr.LDA. This gives the output as sets of words. Each set has words that are related to each other. These sets of words are labeled as different topics. Mr.LDA model approach is used to organizing, summarize large corpus and retrieve topics and words.
Step 3: Conversion into Ontology
 Protégé tool is used to create a terminological ontology using Mr.LDA topic modeling which assigned a set of words under specific topic as shown in Fig. 10 and Fig. 11. For example, the word "according" is detected under topic-19 as shown in Fig. 11. Also, point to be noted in Fig. 10 and 11, one topic is not directly connected to another topic as Mr.LDA is unable to deal with CTM.

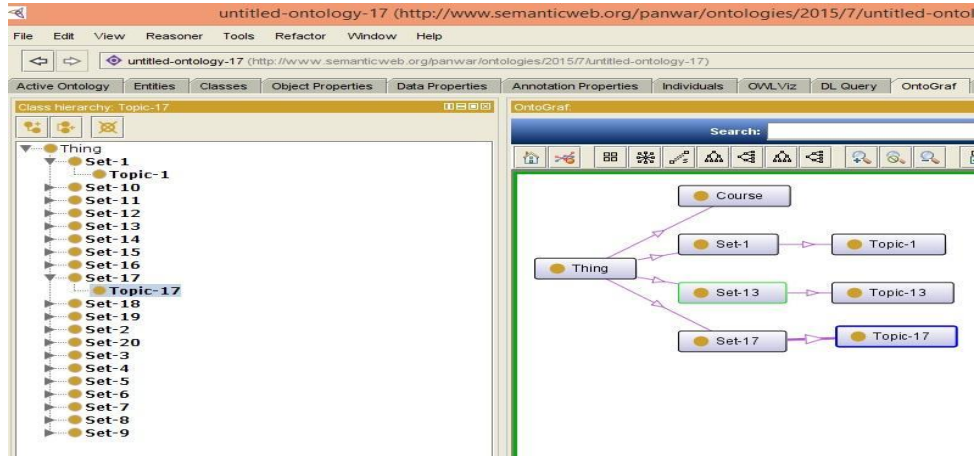

Fig. 10. Protégé tool

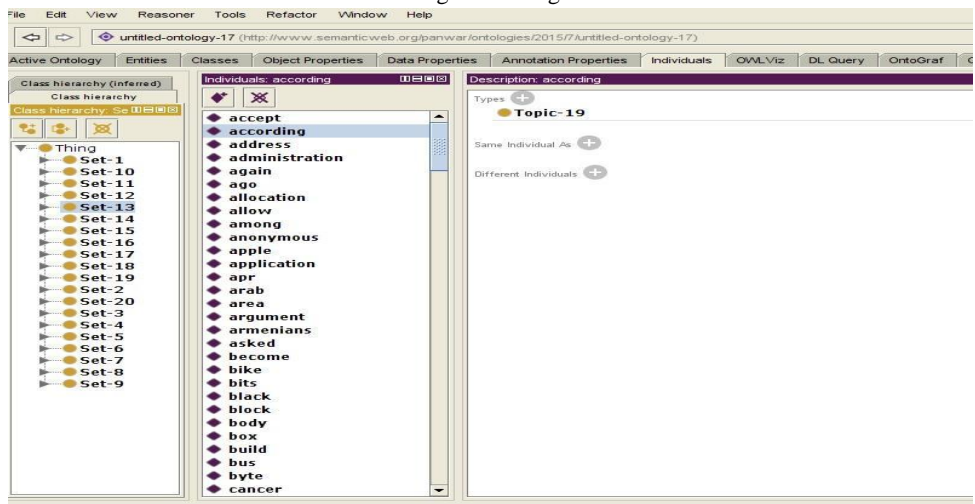

Fig. 11. Terminology Ontology Learning using Mr.LDA model

Step 4: Retrieving results for user queries

The query entered by the user is now processed and its specific topic is identified as shown in Fig. 12. The query is processed depending on the type of query, the particular domain ontologies and accordingly corresponding topics & words (or sentence) will be fetched. For example in Fig. 12, the word "allow" is present in user's query which is detected under topic-9 and topic-2. It might be possible that one word can belong to two or more topics. Similarly, we can perform Topic and Words Detection using SPARQL query on ontologies. Ontologies are used for Topic and Words Detection for Semantic Web (Web 3.0) applications. Terminology ontologies are built to allow the semantic search, provide machine-readable content and managed content and thus, save searching time also cost for user's query.

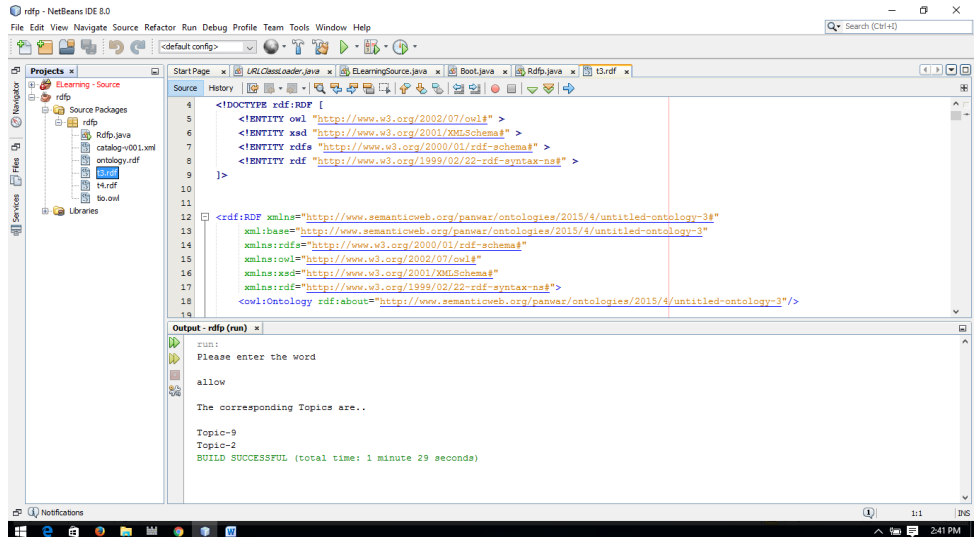

Fig. 12. Retrieving results for user queries

#### 4.2.2 *Naive Bayes Classifier*

Naive Bayes methods are a set of supervised learning algorithms based on applying Bayes' theorem with the "Naive" assumption of independence between every pair of features. In spite of apparently over-simplified assumptions of Naive Bayes, the classifiers have worked quite well in many real-world situations, famously document classification and spam filtering. It requires a small amount of training data to estimate the necessary parameters. Naive Bayes learners and classifiers can be extremely fast compared to more sophisticated methods. The decoupling of the class conditional feature distributions means that each distribution can be independently estimated as one-dimensional distribution. This, in turn, helps to alleviate problems stemming from the curse of dimensionality (Zhang, 2004).

#### 4.2.3 *LDA Model Assumes*

LDA is defined by the statistical assumptions it makes about the corpus. One active area of topic modeling research is how to relax and extend these assumptions to uncover more sophisticated structure in the texts.
- The "Bag-of-Words (BoW)" assumption that the order of the words in the document does not matter.
- The order of documents does not matter.
- The number of topics is assumed known and fixed for LDA.
- LDA algorithm's assumption defines a topic to be a distribution over a fixed vocabulary. For example, the "genetics" topic has words about "genetics" with high probability and the evolutionary biology topic has words about evolutionary biology with high probability. It is assumed that these topics are specified before any data has been generated.

### 4.2.4 LIMITATION

The topics discovered by LDA capture correlations between words, but LDA does not explicitly model correlations among topics. This limitation arises because of the topic proportions in each document sample comes from a single Dirichlet distribution. As a result, LDA has difficulty modeling data in which some topics co-occur more frequently than others. However, topic correlations are common in real-world text data, and ignoring these correlations limits LDA's ability to predict new data with high likelihood. Ignoring topic correlations also hamper the LDA's ability to discover a large number of fine-grained, tightly-coherent topics. Because LDA can combine arbitrary sets of topics, LDA is reluctant to form a highly specific topic for which some combination would be "nonsensical". To resolve this issue Pachinko Allocation Model (PAM), which captures arbitrary, nested, and possibly sparse co-relations between topics using a Directed Acyclic Graph (DAG) is used (Li and McCallum, 2006).

### 4.3 Experimental Evaluation

The effectiveness of the proposed Ontology build by topic modeling using LSI & SVD and Mr.LDA by users is evaluated using 20 Newsgroup dataset as shown in Table 2.

**Table 2**

| Dataset (20 Newsgroup) | |
|---|---|
| comp.graphics | (37 documents), |
| comp.os.ms-windows.misc | (45 documents), |
| comp.sys.ibm.pc.hardware | (40 documents), |
| comp.sys.mac.hardware | (30 documents), |
| comp.windows.x | (75 documents), |
| rec.autos | (48 documents), |
| rec.motorcycles | (32 documents), |
| rec.sport | (40 documents), |
| rec.sport.hockey | (32 documents), |
| sci.crypt | (30 documents), |
| sci.electronics | (54 documents), |
| sci.med | (35 documents), |
| sci.space | (40 documents), |
| talk.politics.misc | (35 documents), |
| talk.politics.guns | (67 documents), |
| talk.politics.mideast | (45 documents), |
| talk.religion.misc | (30 documents), |
| alt.atheism | (30 documents), |
| misc.forsale | (40 documents), |
| soc.religion.christian | (30 documents). |

### 4.3.1 Approach 1: Semantic Ontology Similarity (edge- count method)

Semantic question answers analysis
- User's query is structural analysis (removing stop words)
- WordNet tool has been used to get synonyms of a word of user's query and generate all possible combinations of synonyms.

- Semantic Ontology Similarity (edge-count method using Eq. (3)) match for user's query.
  
  $$S_t(t1; t2) = (e^{xd} - 1) = (e^{xd} + e^{ys} - 2) \qquad (3)$$
  
  where d = depth of the tree, S= shorted path length, x and y are smoothing factors and St (t1; t2) represent similarity value in hierarchical ontology tree.
- The graphical diagram (Fig. 13) is a pictorial representation where the X-axis represents "Ontology Similarity" and Y-axis represent "Keywords".
- Fuzzy Co-clustering & Fuzzy Scale is used to prioritize the retrieved answer.
- Fuzzy Co-clustering to retrieve answers by using Semantic Ontology Similarity was used. Due to various features of Fuzzy Co-clustering can deal with problems like Dimensionality Reduction, Interpretability, accuracy, vague and uncertain query terms.
- To retrieve the semantic answer Fuzzy Scale was used where Fuzzy type-1 (to deal with the uncertainty of document) & Fuzzy type-2 (to deal with the uncertainty of word) (Rani et al., 2014).

For example, a string having synonyms is given as input rather than directly applying a string match: "Passenger claims the airport authority for the their rude behaviour".

After pre-processing and tokenizing the user's query if the search is unable to retrieve the same string ("claim"), then the synonym of the query string from the first set, second set, third set, etc., are analyzed and each set has ten terms. For example the "claim" word in first set (synonyms set is prepared using Google search engine) having following synonyms example: aver ($\mu=0.51$), avow ($\mu=0.52$), affirm ($\mu=0.53$), hold ($\mu=0.54$), state ($\mu=0.55$), maintain ($\mu=0.56$), profess ($\mu=0.57$), declare ($\mu=0.58$), assert ($\mu=0.59$).

$$\text{Score } (S_1) = (A + A\tilde{}/N). \qquad (4)$$

Where "A" represent Membership of document, "A~" represent Membership of words and Number of document (N). Here, Score ($S_1$) is calculated using Eq. (4):

Score ($S_1$) = (0.5 + 0.59)/2 = 0.795.

Ontology similarity is useful to retrieve semantic relations in response to the user's query. In most of the cases, the search results remain the same for both ontologies. Using Mr.LDA and LSI & SVD change can be noticed in terms of time complexity. As to build the ontology using Mr.LDA is fast compared to LSI & SVD.

### 4.3.2 *Approach 2: Ranking of ontologies*

The rank of ontologies with respect to user's query is evaluated on the base of following criteria: Class Match Measure (CMM), Betweenness Measure (BEM), Density measure (DEM) and Semantic Similarity Measure (SMM) (Colace et al., 2014).

CMM - User's query terms are search in Terminology ontology if found exactly or partially then score is assigned. The score is dependent on the "coverage" of user's query term in terminology ontology.

DEM - DEM is concerned with rich knowledge representation of types, properties, relationship types and interrelationship of concepts for the specific topic. Various research fields like computer science, information science etc. use rich knowledge representation for logical inferences.

SMM - SSM is concerned with the minimum number of taxonomic links that connects one concept to another. User's query term is match with concepts of terminology ontology using Rada Distance Measure.

BEM - BEM is calculated for the concepts node which occurs frequently in the many shortest paths. The frequently occurring central concept node will be assigned a higher score in terminology ontology.

- Let O is the set of ontologies to rank and M= {M[1],…..M[i], M[4]} = { CMM, DEM, SSM, BEM} (Alani et al., 2006). For calculating Score (using Eq. (4)) weights set have been used for our experiment are 0.1, 0.2, 0.3 and 0.4 for the BEM, DEM, CMM and SSM measures respectively as shown in Table 3.
- Score calculated ($S_2$) =
$$\text{Score } (o \in O) = \sum_{i=1}^{4} W_i \frac{M[i]}{\max_{1 \leq j \leq |o|} M[j]} \quad (4)$$
- The graphical diagram (Fig. 14) is a pictorial representation where the X-axis represents "Measure value" and Y-axis represents "CMM, BEM, DEM, and SMM".

**Table 3**
Ranking of ontologies.

| Ontology | CMM | DEM | SSM | BEM | Score | Rank |
|---|---|---|---|---|---|---|
| Ontology using LSI & SVD | 0.235 | 0.5 | 0.275 | 0.632 | 0.3437 | 2 |
| Ontology using Mr.LDA | 0.342 | 0.486 | 0.349 | 0.822 | 0.4216 | 1 |

Semantic Ontology Similarity (Approach 1) and Ranking of Ontologies (Approach 2) lead to the conclusion that Ontology build using Mr.LDA is better than the ontology build using LSI & SVD. Approach 1 and Approach 2 is used to calculate Score ($S_1$) and Score ($S_2$) respectively. To show experimental comparison we used same dataset (20 newsgroups) for topic modeling (LSI & SVD and Mr.LDA). But for the large dataset (hundreds of GB's, or TB's of data) Mr.LDA utilizes the entire given cluster of machines whereas LSI &SVD utilize one machine to obtain topics and words distribution. Mr.LDA expected to run fast due to the parallel nature of MapReduce.

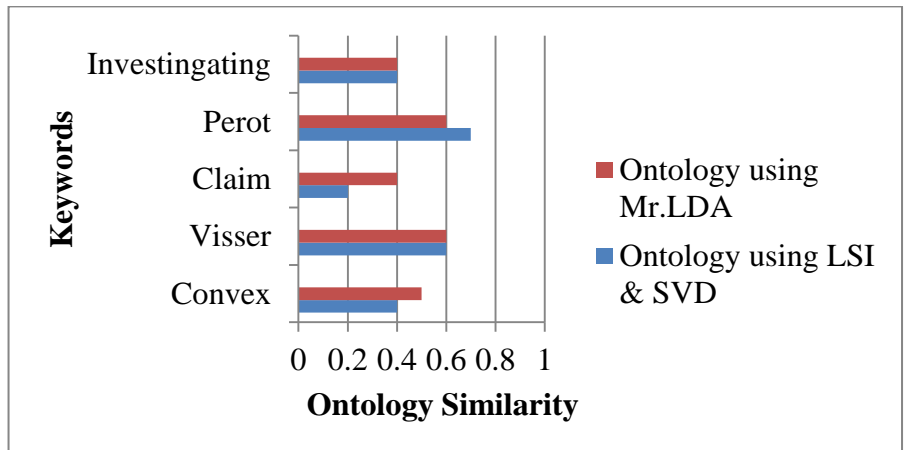

Fig. 13. Ontology Similarity

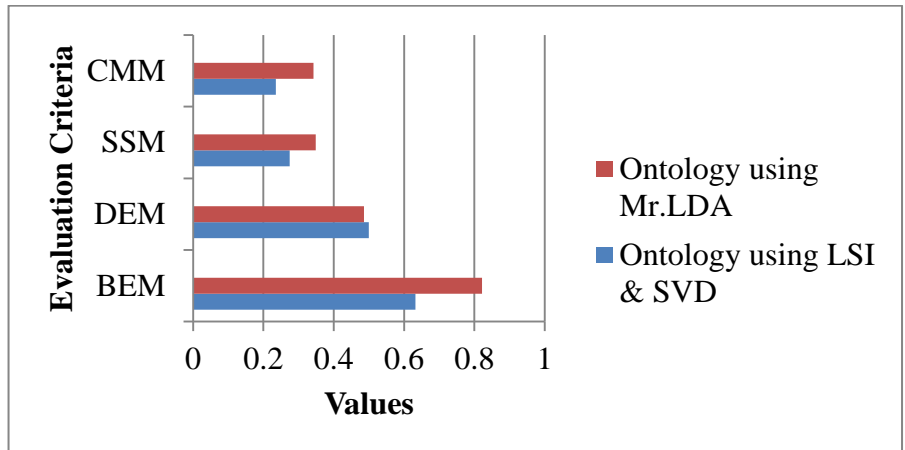

Fig. 14. Evaluation Criteria

**5. Conclusion and Future work**

In this paper, topic modeling algorithms were explored, namely LSI & SVD and Mr.LDA for Ontology Learning (OL). The study and experimental result give enough proof of the effectiveness of using Mr.LDA topic modeling for OL. Experimental results in the paper demonstrate the effectiveness of the proposed system in term of building richer topic-specific knowledge and semantic retrieval. Terminology ontology building is a preliminary step for semantic-based query (Topics and Words Detection) optimization for knowledge management.

Future work will focus on OL for the specific domain that can allow semantically accurate retrieval without human intervention. For OL, we can consider other topic modeling example Hierarchical Dirichlet Process (HDP) and Pachinko Allocation Model (PAM) modeling algorithm instead of Mr.LDA as it suffers from Correlation Topic Modeling (CTM). The fully automatic OL can provide a solution to represent knowledge in the various fields example: Topic Detection and Tracking, Knowledge Engineering, Bio-informatics, Artificial Intelligence (AI), Natural Language Processing (NLP), E-commerce, Education and in new emerging fields like the Semantic Web for e-Learning (SWEL). Specifically, for e-learning field, Domain ontology Construction (DOC) encodes the knowledge and makes it reusable for learners. Information Retrieval using agents on the knowledge encode on ontology for Learning Path Generation (LPG), Object Recommendation (OR), Personalization of Content (POC), and thus can improve semantic web education learning (SWEL) is possible with the proposed approach.